\documentclass[aps,prl,preprint,groupaddress]{revtex4}

\usepackage{bbm}
\usepackage{graphicx}
\usepackage{subfigure}
\usepackage{amsmath}
\usepackage{booktabs}

\begin{document}

\title{Strange memory effect of low-field microwave absorption in copper-substituted lead apatite}

\author{Jicheng Liu$^{1,\#}$, Chenao He$^{2,\#}$, Weijie Huang$^{3,\#}$, Zhihao Zhen$^{3}$, Guanhua Chen$^{3}$, Tianyong Luo$^{5}$, Xianfeng Qiao$^{2,4}$\footnote{\url{msxqiao@scut.edu.cn}}, Yao Yao$^{2,3}$\footnote{\url{yaoyao2016@scut.edu.cn}}, and Dongge Ma$^{2,4}$}

\address{$^1$ School of Minerals Processing and Bioengineering, Central South University, Changsha 410083, China\\
$^2$ State Key Laboratory of Luminescent Materials and Devices, South China University of Technology, Guangzhou 510640, China\\
$^3$ Department of Physics, South China University of Technology, Guangzhou 510640, China\\
$^4$ Guangdong Provincial Key Laboratory of Luminescence from Molecular Aggregates, Guangdong-Hong Kong-Macao Joint Laboratory of Optoelectronic and Magnetic Functional Materials, South China University of Technology, Guangzhou 510640, China\\
$^5$ School of Mechanical and Electrical Engineering, University of Electronic Science and Technology of China, Chengdu 611731, China}

\date{\today}

\begin{abstract}
We observe a considerable hysteresis effect of low-field microwave absorption (LFMA) in copper-substituted lead apatite. By continuously rotating samples under external magnetic field, this effect is diminished which can not be renewed by a strong magnetic field but will be spontaneously recovered after two days, indicating its glassy features and excluding possibility of any ferromagnetism. The intensity of LFMA is found to sharply decrease at around 250K, suggesting a phase transition takes place. A lattice gauge model is then employed to assign these effects to the transition between superconducting Meissner phase and vortex glass, and the slow dynamics wherein is calculated as well.
\end{abstract}

\maketitle

Researches on applicable superconductors have attracted interests over 100 years \cite{17,18,19}. The critical temperature ($T_{\rm c}$) of superconducting transition always serves as the essential parameter for the practical application as the highest superconducting temperature under ambient pressure is still below 150K \cite{25,20,21}. With the development of experimental technology and material synthesis, $T_{\rm c}$ with high pressure has already achieved above 250K, but the indispensable pressure on the order of 100 GPa acts as a serious obstacle \cite{22,23}. Very recently, Lee et al. claimed they successfully synthesized a new type of superconductor at room temperature and ambient pressure, i.e. copper-substituted lead apatite (CSLA) ${\rm Pb_{10-x}Cu_x(PO_4)_6O}$, which is also called LK-99 \cite{2,3}. To verify this astonishing finding, several independent groups have followed the procedure of synthesis and in terms of X-ray diffraction features, the claimed structure of CSLA can be basically proven \cite{4,5,6,12,Wang2023,Guo2023,cutedai}. Up to date, possible Meissner effect and zero resistance at room temperature have not been reported. In addition, some first-principles calculations with predicted crystal structure also provide theoretical supports on the possible superconductivity \cite{7,8,9,10,11,prbPrincetonDFT,prbViennaDFT,prbyuechangmingDFT}.

As suggested by Lee et al., the structure of CSLA possesses two circles: The outer circle serves as a shield to protect the inner one which forms a quasi-one-dimensional (1D) conducting channel  \cite{2,3}. The essential idea is to substitute the outer lead atoms with copper to shrink the whole structure. This 1D superconductivity model can be well applied to explain the anisotropic levitation posting on the social media and thus greatly inspires us to uncover the possible 1D strongly-correlated mechanism with a magnetic flux. Previously, we have reported that the cuprate radicals in CSLA hold sufficiently long coherence time to be quantum manipulated \cite{liu2023longcoherence}, which yields a useful hint for a successful synthesis. So far, only the powder of mixture has manifested possible superconducting features, so normal electric and magnetic measurements are not available in the current stage. Learnt from the research history of other superconducting materials, such as Y-Ba-Cu-O \cite{1987prblowfield,1987JP_CYBCO,1987prbYBCO,1990prbYBCO,1991prbYBCO}, alkali-metal-doped fullerene \cite{1992jpcfullerene}, magnesium diboride \cite{2001MgB2,2003PCMgB}, and iron pnictides \cite{2017FeAs}, the detection of microwave absorption turns out to be an appropriate approach to determine whether there is superconducting phase in the mixtures, which motivates the main subject of the present work.

We follow the procedure of Lee et al. to synthesize our CSLA samples, with $1:1$ mole ratio of Pb$_2$(SO$_4$)O and Cu$_{3}$P \cite{2,3,liu2023longcoherence}. The samples are then measured with cw microwave absorption spectroscopy on an EPR spectrometer (Bruker ELEXSYS E580) operated at the X-band (9.667~GHz) and outfitted with a dielectric resonator (ER-4118X-MD5). The microwave power is 4.743~mW, and the modulation amplitude is 1~Gauss at 100~kHz. The magnetic field was corrected by using a BDPA standard (Bruker E3005313) with $g=2.0026$. Low temperature environment is realized by an Oxford Instruments CF935 continuous-flow cryostat using liquid nitrogen. The temperature is controlled by an Oxford ITC4 temperature controller with accuracy of $\pm$0.1~K.

\begin{figure*}[t]
    \centering
    \includegraphics[width=1.0\linewidth]{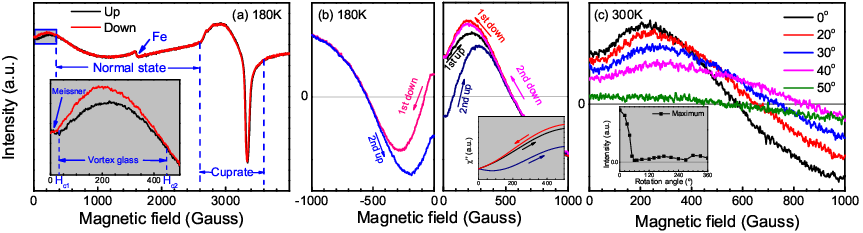}
    \caption{(a) Microwave absorption spectra at 180~K, with three remarkable signals: Cuprate radicals from 2600 to 3600~Gauss, a broad negative peak as called normal state from 500 to 2600~Gauss, a positive derivative LFMA below 500~Gauss. The small peak at around 1700~Gauss is from iron in quartz tube. Inset displays the amplified peaks of derivative LFMA with a small plateau named Meissner and a wide peak named vortex glass. Sweeping the magnetic field up and down, a clear hysteresis effect is observed for the LFMA. The turning point and the bifurcation point are recognized as lower and upper critical field $H_{\rm c1}$ and $H_{\rm c2}$, respectively. (b) The hysteresis curve of derivative LFMA in the first and second sweeps. The curves of negative field are obtained by rotating the sample to 180$^{\circ}$ and reversing the signs of both signal and magnetic field. Inset shows the integrated spectra, namely the imaginary part of ac susceptibility $\chi''$. (c) Derivative LFMA with various rotation angles at room temperature. As the sample is powder, the initial angle 0$^{\circ}$ is defined arbitrarily. Inset plots the maximum intensity versus the rotation angle.}
    \label{fig1}
\end{figure*}

As reported previously \cite{liu2023longcoherence}, the cuprate radicals contribute the most visible response to the X-band microwave under external magnetic field of around 3350~Gauss, as illustrated in Fig.~\ref{fig1}(a). This strong paramagnetic signal would conceal others in normal dc magnetic measurements but can be well distinguished from others indicating advantages of utilizing microwave technique. Significantly, from 0 -- 2600~Gauss, there is a super broad absorption signal that is interesting to be investigated, except a small kink from irons in the quartz tube. We divide this region into three phases. A small plateau below 30~Gauss, a positive signal (30 -- 500~Gauss) and a negative signal (500 -- 2600~Gauss) will be called Meissner, vortex glass and normal state \cite{1994rmp}, as discussed in details below.

Most superconductors have got the low-field microwave absorption (LFMA) due to the presence of superconducting gap and the relevant superconducting vortices as excited states \cite{1995superconducting_review}. More importantly, the derivative LFMA of superconductors is positively dependent of the magnetic field as the vortices are more induced under higher field. As a comparison, although the soft magnetism is also active under low field, the precession of spin moments will be suppressed so that the derivative LFMA of magnetic materials is normally negative. The sign of LFMA can be always corrected by the signal of radicals in our measurements. In our cases, the signals below 500~Gauss are all positive, implying the presence of superconductivity.

We then sweep the magnetic field forward and backward and observe a prominent hysteresis effect below 450~Gauss, which is independent of the sweep rate. Above this field, the hysteresis is completely absent, excluding the possibility that the positive LFMA and the negative high-field signal together constitute a ferromagnetic resonance (FMR) signal. We guess the negative one refers to a magnetoresistance effect in the normal state. The first turning point and the bifurcation point can be realized as the lower and upper critical field $H_{\rm c1}$ and $H_{\rm c2}$, which in this case are 30 and 450~Gauss, respectively.

In order to display an entire hysteresis curve, one has to reverse the orientation of magnetic field, but due to the instrument limitation, we are solely able to rotate the samples to 180$^{\circ}$ and then reverse the signs of both signals and magnetic fields, as shown in Fig.~\ref{fig1}(b). A pretty hysteresis curve is found, by which one can see as reversing the orientation the signal is almost continued. The values are not explicitly smoothed partly because the baselines are not deducted. If we understand the absorption of microwave is enabled by the generation of vortices, they have not got enough time to relax giving rise to this hysteresis. It is worth noting that, the EPR signal is nothing but the derivative of the imaginary part of ac magnetic susceptibility, namely $\frac{d\chi''}{dH}$, and this hysteresis actually points out the feature of relevant excited states with regard to the dc magnetization curves. We thus integrate the signal and plot the imaginary ac susceptibility $\chi''$, from which a closely linear relationship with dc magnetic field is observed \cite{2022ma15031079}.

The hysteresis effect suggests us to further examine other orientations of magnetic field. We then rotate the sample under zero field from an initial angle, defined as 0$^{\circ}$ for convenience, and at every 10$^{\circ}$ the magnetic field is swept from 0 to 5000~Gauss. As displayed in Fig.~\ref{fig1}(c), it is found that following rotation, the LFMA is rapidly diminished until almost disappears, implying the absorption of microwave is saturated. Afterward, the signals can not be renewed in a short duration, regardless of whether we continue to rotate the sample to its initial angle or enhance the magnetic field to 9600~Gauss. This strange memory effect of magnetic field orientation strongly eliminates the possible contribution of any ferromagnetism, which can not be killed by magnetic field. After about two days of rest in the ambient circumstance, the signal of samples is spontaneously recovered. We thus realize, the memory effect figures out the slow dynamics of vortex creep in the glass phase \cite{1987prblowfield}.

\begin{figure*}[t]
    \centering
    \includegraphics[width=0.5\linewidth]{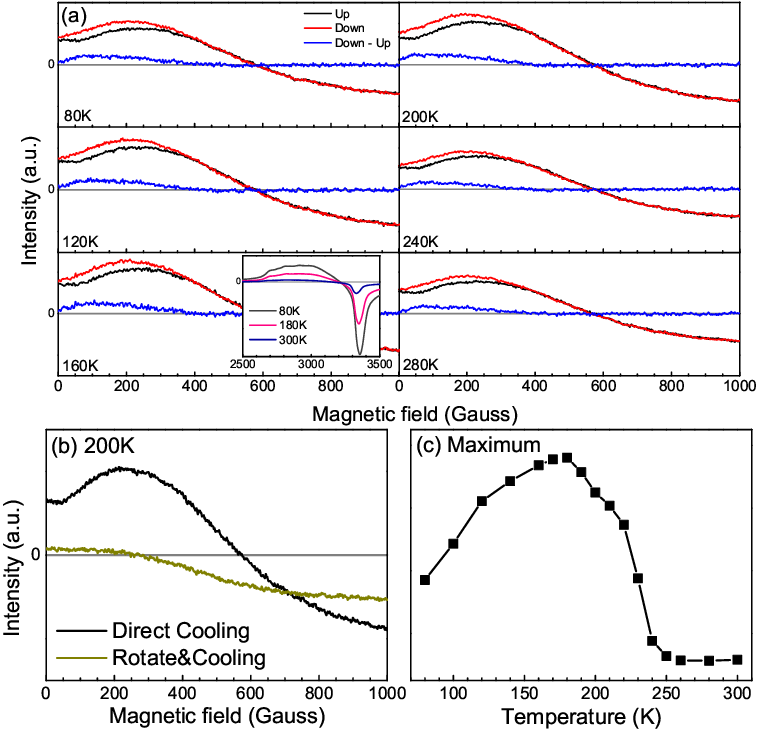}
    \caption{(a) Derivative LFMA signals of magnetic field sweep up, down and down minus up at six temperatures from 80 -- 280~K. To be comparisons, the inset shows the signals of cuprate radicals at three temperatures, respectively. (b) Derivative LFMA signals at 200~K by directly cooling the relaxed sample and first rotating and then cooling. (c) The maximum intensity of derivative LFMA versus temperature. The transition point is around 250~K.}
    \label{fig2}
\end{figure*}

The temperature dependence of derivative LFMA is displayed in Fig.~\ref{fig2}(a). The hysteresis effect is visible at all temperatures, and peak positions do almost not change. This weak temperature dependence does also not support magnetic response, since following temperature increasing the FMR would become sharper and closer to the EPR peak. As comparisons, we also plot the relevant EPR spectra at different temperatures, which normally exhibits significant decrease with temperature increasing. In Fig.~\ref{fig2}(b), the comparing results of ``direct cooling" and ``rotate\&cooling" are illustrated. The former is to directly cool the sample without any initial magnetization, and the latter is to firstly rotate the samples in the magnetic field for saturated absorption and then cool to 200~K. It is found that, after saturated absorption the LFMA disappears even at low temperature, while the high-field negative signals are recovered in a portion, again demonstrating they stem from different mechanisms. The maximum intensity of derivative LFMA versus temperature is displayed in Fig.~\ref{fig2}(c). Following temperature increasing, it increases first and then sharply decreases from 190~K, suggesting a phase transition takes place. The turning point is around 250~K, which can be recognized as the critical temperature $T_{\rm c}$.

Above experimental results together indicate the primary features of CSLA: positive LFMA, hysteresis effect during magnetic field sweep, saturated absorption while rotation with strangely long memory effect, weak temperature dependence with a phase transition. We therefore appoint the most possible mechanism to the superconducting vortices. The low-field absorption of microwave power with the assistance of dc magnetic field points to the small superconducting gap, and the relevant metastable excited states emerge to be vortices. The vortex creep and relaxation have got glassy slow dynamics to yield the memory effect in both field sweep and rotation. Since the samples are in powder phase, the random orientation of vortices in a quasi-1D lattice makes them solely respond to the magnetic field with proper orientation. Magnetic vortex can not be killed by the magnetic field, so the long-standing vortex state can only be thought of stemming from superconductivity.

\begin{figure*}[t]
    \centering
    \includegraphics[width=0.5\linewidth]{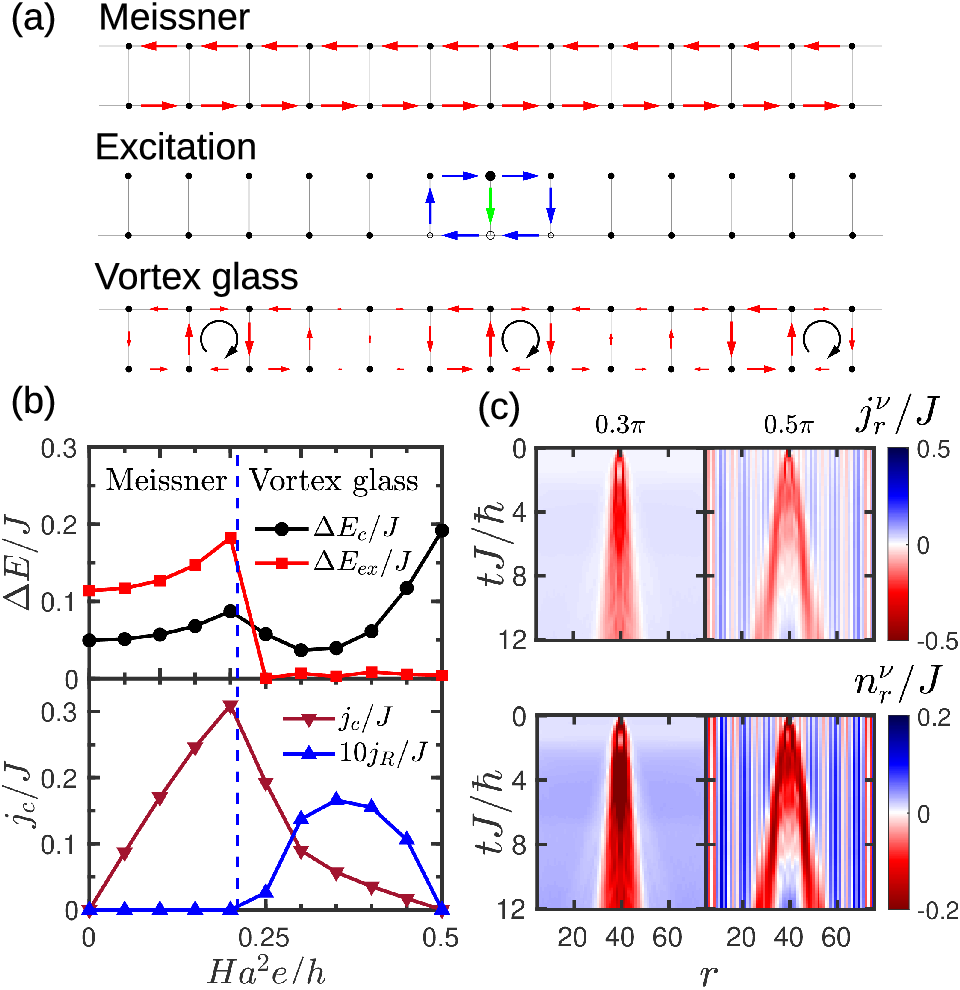}
    \caption{(a) Typical current patterns for the Meissner and vortex glass phase, as well as the current excitation. The length of red arrows and the size of dots label the magnitude of current and density, respectively. The rung current excitation is sketched by green arrow, and deviations of currents from the background chiral currents are represented by blue arrows. (b) The charge and excitation gap and chiral current are plotted against the magnetic flux to illustrate the Meissner and vortex glass phase at half filling. The rung current $j_{R}$ is averaged over $\left | \langle j_r^{\perp} \rangle \right | $ on the central half of the lattice. (c) Time evolution of local observables after current excitations with two $\phi$'s: $\phi=0.3\pi$ for Meissner and $\phi=0.5\pi$ for vortex glass.
    }
    \label{fig3}
\end{figure*}

In order to determine whether the two phases Meissner and vortex glass can be tuned by magnetic field, we consider a quasi-1D flux ladder model in an artificial gauge field \cite{orignac2001meissner,roux2007diamagnetism, piraud2015vortex, greschner2016symmetrybroken}. Due to the powder phase, the realistic orientations of the 1D ladder are regarded as random. The Hamiltonian is written as
\begin{equation}
    \begin{aligned}
    H= & -J \sum_{r=1}^{L-1} \sum_{\ell=1}^2\left(a_{\ell, r}^{\dagger} a_{\ell, r+1}+\text { H.c. }\right)\\
       &-J_{\perp} \sum_{r=1}^{L} \left( e^{-i r \phi}a_{1, r}^{\dagger} a_{2, r}+\text { H.c. }\right),
    \end{aligned}
    \label{eq:1}
\end{equation}
which is on a ladder lattice with $L$ rungs as sketched in Fig.~\ref{fig3}(a). Herein, $a_{\ell, r}^{\dagger} (a_{\ell, r})$ creates (annihilates) a local Cooper pair on the lower $(\ell=1)$ or the upper $(\ell=2)$ leg of the $r$-th rung; $J$ and $J_{\perp}$ label the nearest-neighbor hopping terms along legs and rungs, respectively. We define the filling factor as $f=N/(2L)$, where $N$ is the total number of Cooper pairs, and consider the limit of hard-core bosons with half filling ($f=0.5$). The flux per ladder plaquette $\phi$ is the essential parameter that is proportional to the magnetic field $H$, namely $\phi = 2\pi H a^{2}e/h$, where the lattice constant $a$ is related to the coherent length of the Cooper pair, $e$ is the electron charge, and $h$ is the Planck constant. For the experimental considerations, a flux of $\phi = 1.0\pi$ corresponds to a magnetic field of $H\sim500$~Gauss when $a\sim200$~nm. In the calculations, we set $J = J_{\perp}= 1$ as the unit of energy.

For the sake of discussing the slow dynamics of vortex creep \cite{1994rmp}, we define a local current operator on legs as $j_{\ell, r}^{\|}=i J\left(a_{\ell,  r}^{\dagger} a_{\ell, r+1 } -a_{\ell, r+1}^{\dagger} a_{\ell, r} \right)$ and on the rung as $j_r^{\perp}=i J_{\perp}\left(e^{-i r \phi} a_{1, r}^{\dagger} a_{2, r}-e^{i r \phi} a_{2, r}^{\dagger} a_{1, r}\right)$. A background chiral current is then defined as $j_{\rm c} = \frac{1}{L-1} \sum_{r=1}^{L-1}  \langle j_{1, r}^{\|} -j_{2, r}^{\|} \rangle$ to characterize the average current circulating the ladder along the legs \cite{orignac2001meissner,roux2007diamagnetism,piraud2015vortex}. The ground state is calculated by using the density matrix renormalization group method \cite{schollwock2011density,itensor}. We simulate the flux ladder up to $L=201$ rungs and the bond dimension is typically up to 1000. The subsequent time evolution is simulated via time evolving block decimation method \cite{White2004}, and the discarded weight is fixed as ${\rm cutoff}=10^{-6}$.

We focus on the transition from the Meissner-Mott insulator state (Meissner) to the gapless vortex-Mott insulator state (vortex glass) by tuning the magnetic field. This phase transition in the flux ladder model has been comprehensively discussed in Ref.~\cite{piraud2015vortex}, so we can easily follow their footsteps. In Fig.~\ref{fig3}(b) we plot the charge gap $\Delta E_{\rm c} = E^{0}_{N+1} + E^{0}_{N-1} - 2E^{0}_{N}$ with $E^{m}_{n}$ being energy of the $m$-th state in the sector of $n$ particles and the excitation gap $\Delta E_{\rm ex} = E^{1}_{N} - E^{0}_{N}$ as a function of magnetic flux $H a^{2}e/h$ to depict the phase transition. The terms Meissner and vortex state account for the characteristic current patterns as also sketched \cite{orignac2001meissner,piraud2015vortex,greschner2016symmetrybroken}. At small flux, there is a Meissner state having vanishing rung currents $\langle j_r^{\perp} \rangle = 0$ but a finite chiral current $j_{\rm c}$. By increasing the flux up to a critical value, the system enters into the vortex glass phase where local currents in the inner rungs arise, namely $\langle j_r^{\perp} \rangle \ne 0$, leading to the formation of vortices which are distributed without any periodicity. While the Meissner state is completely gapped, the vortex glass exhibits one gapless mode such that the excitation gap $\Delta E_{\rm ex}$ vanishes while the charge gap $\Delta E_{c}>0$.

To study the vortex creep, the single rung in the middle is excited to generate vortex current at time $t=0$ by applying a rung current operator without phase factor to the ground states, i.e., $i \left( a_{1, L/2}^{\dagger} a_{2, L/2} - a_{2, L/2}^{\dagger} a_{1, L/2}\right) \left | \psi_{0}  \right \rangle $ \cite{huang2023spatial}. The excitation leads to the changes of the local current and particle density which generate vortices slowly moving outwards from the middle rung. The deviations from mean values fingerprint the creeping vortices, so we define $j^{\nu}_{r} = \left \langle j^{\|}_{1,r} - j^{\|}_{2,r} -j_{\rm c} \right \rangle$ and $ n_{r}^{\nu} = \left \langle n_{1,r}+n_{2,r}-1\right \rangle$. In Fig.\ref{fig3}(c), we simulate the Meissner state ($\phi=0.3\pi$) and vortex glass ($\phi=0.5\pi$) for comparisons. For the Meissner state, the vortices are localized in the middle rung and broadens with time, while in the vortex glass the vortices separate into two branches and individually propagate outwards manifesting a clear picture of vortex creep.

In summary, we have found significant hysteresis and memory effect of LFMA in samples of CSLA. The effect is sufficiently robust in magnetic field sweep and rotation and will lose memory in a long duration. The temperature dependence of LFMA intensity exhibits a phase transition at 250~K. The phase diagram of superconducting Meissner and vortex glass is then calculated in the framework of lattice gauge model. In the near future, we will continue to improve the quality of samples to realize full levitation and magnetic flux pinning by increasing active components. The application of a microwave power repository will be considered as well.

\section{Acknowledgments}

The authors gratefully acknowledge support from the National Natural Science Foundation of China (Grant Nos.~11974118, 12374107, 61975057, 21788102 and 51527804), the National Key R\&D Program of China (2020YFA0714604), the Foundation of Guangdong Province (2019B121205002), the Key Research and Development Project of Guangdong Province (Grant No.~2020B0303300001), and the Proof-of-concept Funds of Fudan ZhangJiang Institute.

\section{Author contributions}

\# These authors equally contributed to this work.

X.Q. and Y.Y. designed and conducted the EPR experiments and analyzed the data with the help of C.H. in SCUT. J.L. synthesized and characterized the CSLA samples with the help of Z.Z. in CSU. W.H. and Y.Y. performed the simulations and theoretical analysis with G.C. in SCUT. Y.Y. wrote the manuscript with the help of other authors. T.L. and D.M. contributed useful inputs to the work.

\section{Competing interests}

The authors declare no competing interests.

\bibliography{LF_v7.bbl}

\end{document}